\documentclass[showpacs,twocolumn]{revtex4}

\usepackage{graphicx}
\usepackage{amsmath}
\usepackage{amssymb}

\begin{document}

\title{Glass transition of hard spheres in high dimensions}
\date{\today }

\begin{abstract}
We have investigated analytically and numerically the liquid-glass transition
of hard spheres for dimensions $d\rightarrow \infty $ in the framework of mode-coupling theory.
The numerical results for the critical collective
and self nonergodicity parameters $f_{c}\left( k;d\right) $ and $%
f_{c}^{\left( s\right) }\left( k;d\right) $ exhibit non-Gaussian $k$%
-dependence even up to $d=800$. $f_{c}^{\left( s\right) }\left( k;d\right) $\ and $f_{c}\left(
k;d\right) $ differ for $k\sim d^{1/2}$, but become identical on a scale $k\sim d$,
which is proven analytically. The critical packing fraction
$\varphi _{c}\left( d\right) \sim d^{2}2^{-d}$ is above the corresponding Kauzmann
packing fraction $\varphi _{K}\left( d\right)$ derived by a small cage expansion.
Its quadratic pre-exponential factor is different from the linear one found earlier.
The numerical values for the exponent parameter and therefore the
critical exponents $a$ and $b$ depend on $d$, even for the largest values of 
$d$.
\end{abstract}

\author{Bernhard Schmid}
\affiliation{Institut f\"ur Physik, Johannes Gutenberg-Universit\"at Mainz, Staudinger
Weg 7, D-55099 Mainz, Germany}
\author{Rolf Schilling}
\affiliation{Institut f\"ur Physik, Johannes Gutenberg-Universit\"at Mainz, Staudinger
Weg 7, D-55099 Mainz, Germany}
\maketitle

\section{INTRODUCTION}

In many situations the analytical treatment of a specific physical problem
simplifies drastically if the spatial dimension $d$ becomes infinite. For
instance, it is well-known that the mean field theory for systems in thermal
equilibrium becomes exact for $d=\infty $. The equation of state of a fluid
can be obtained from a virial expansion. For the fluid of \textit{hard
spheres} it has been shown that for such packing fractions $\varphi $ for
which the second virial term (which is proportional to $\varphi $) is finite
or at most algebraically increasing with $d$, the third and higher order
virial terms vanish exponentially fast in the limit $d\rightarrow \infty $ 
\cite{HRW85,WRH87}.

Hard sphere systems are ideal systems to study not only equilibrium
properties, but also the liquid-glass transition and glassy dynamics. The
mode-coupling theory (MCT) \cite{GOE09} is a microscopic theory of an ideal
glass transition. Knowledge of the static density correlators allows to
calculate the long time relaxation of a supercooled or supercompressed fluid
and to locate the glass transition point at which the ergodic behavior in
the fluid phase changes discontinuously into a nonergodic one. The
experimental results for colloidal fluids, which can be modelled by hard
spheres, exhibit agreement with the corresponding MCT result after the
transient regime over several decades in time within ten percent precision 
\cite{MU93, SP05}.

Properties of equilibrium phase transitions, e.g. the critical exponents at
a second order phase transition, depend strongly on the spatial
dimensionality. This has motivated the investigation of the glass
transition for $d=2$ \cite{BAY07,HAJ09,K0B09} and $d=3,4$ \cite{K0B09,ER09,CIMM09}.
The most important approximation of MCT is the factorization of the memory
kernel \cite{GOE09}. This kernel is a \textit{time-dependent} four-point
correlator of the density modes $\rho(\vec{k})$ which is approximated
by a product of time dependent two-point correlators. This factorization
resembles the mean field approximation replacing a static two-point
correlator, e.g. the spin-spin correlator $\langle S_iS_j\rangle$ for an
Ising model, by a product of the order parameter, e.g. the magnetization
$\langle S_i\rangle$ in case of the Ising model. Based on this analogy,
MCT has been interpreted as a mean field theory with the two-point density correlator
as an order parameter \cite{BIR04}, where spatial fluctuations of the correlation
between the pair densities $\rho(\vec{r},t)\rho(\vec{r}+\vec{\delta},t+\tau)$
and $\rho(\vec{r}^{\,\prime},t^\prime)\rho(\vec{r}^{\,\prime}+\vec{\delta},t^\prime+\tau)$ are neglected.
In a next step, these spatial fluctuations are taken into account. Finally it is shown
that the upper critical dimension where the spatial fluctuations do not influence
the critical behaviour is $d_c=6$ \cite{BIR04}, for systems
without and $d_c=8$ \cite{BB07,FPRR10} with
conserved quantities. This implies that the square root
singularity of the nonergodicity parameter and the relation between
the exponent parameter $\lambda (d)$ and the ``critical'' exponents
$a(d)$, $b(d)$ \cite{GOE09} are universal above $d_c$ \cite{ABB09}.
However, the exponent parameter $\lambda(d)$ itself being determined by the
static structure factor at the glass transition singularity, depends
on $d$. The interpretation of MCT as a mean field model challenges
the investigation of MCT with full $k$-dependence for $d\rightarrow \infty$. As
already mentioned above, analytical calculations simplify for
$d\rightarrow \infty $, e.g. the leading order term of the static and
direct correlation function for hard spheres are known and become dominant
(see below). Consequently we will focus on the MCT glass transition of
hard spheres in high dimensions.

Let us shortly review what is already
known for hard spheres and $d\rightarrow \infty$. Taking for the direct
correlation function the leading order of a virial expansion (see below),
using the Vineyard approximation \cite{BOO80} for the normalized collective
nonergodicity parameters $f(k;d)$, i.e. it is $f(k;d)\approx f^{(s)}(k;d)$,
and assuming the nonergodicity parameters $f^{(s)}(k;d)$ of the self
correlator to be \textit{Gaussian} in $k$ with width $\alpha $, a self
consistency equation for $\alpha $ follows from MCT. As critical packing
fraction for the glass transition it has been found \cite{KW87}%
\begin{equation}
\varphi _{c}^{KW}\left( d\right) \cong \sqrt{2\pi e}~d2^{-d}, ~~~d\rightarrow
\infty .  \label{eq2}
\end{equation}%

The replica theory for the structural glass transition \cite{MP} is another
microscopic theory. It allows to calculate the Kauzmann temperature $T_K$ or
the corresponding packing fraction $\varphi_K$ at which the configurational
entropy per particle vanishes. Applied to hard spheres in high dimensions
and performing a \textit{small cage expansion} it is found \cite{PAR06}
\begin{equation}
\varphi _{K}\left( d\right) \cong d\ln \left( d/2\right) 2^{-d}, ~~~d\rightarrow
\infty .  \label{eq3}
\end{equation}%

Our main motivation is to explore the MCT scenario for $d\rightarrow \infty$,
i.e. we want to investigate whether the $A_2$ singularity \cite{GOE09} of MCT
in $d=3$ survives for $d\rightarrow \infty$.
Furthermore, we want to check whether the critical
nonergodicity parameters $f_c(k;d)$ and $f_c^{(s)}(k;d)$ of the collective and
self correlators, respectively, are Gaussian and become equal for
$d\rightarrow \infty$ and whether the critical packing fraction $\varphi_c(d)$
coincides with $\varphi^{KW}_c(d)$ and if not how it relates to the Kauzmann
value $\varphi_K(d)$. In a first step we have solved
numerically the MCT equations for the \textit{collective} and \textit{self}
nonergodicity parameters $f(k;d)$ and $f^{(s)}(k;d)$, respectively up to $d=800$. Inspired by the
numerical solution we have investigated in a second step the corresponding
equations analytically. The outline of our paper is as follows. The next
section presents the MCT equations in arbitrary dimensions $d$ and their
numerical solution for the nonergodicity parameters of the collective and self correlator.
Based on these numerical results we present in the third section an
analytic investigation of the MCT equations for hard spheres for $d\rightarrow \infty$.
The final section IV contains a summary and conclusions.

\section{MCT EQUATIONS AND NUMERICAL SOLUTION}

\subsection{MCT equations}

We consider $N$ hyperspheres with diameter $\sigma $ in a $d$-dimensional
box with volume $V$. The number density is $n=N/V$ and the packing fraction:%
\begin{equation}
\varphi =nV_{d}\left( \sigma /2\right)  \label{eq4}
\end{equation}%
with%
\begin{equation}
V_{d}\left( R\right) =\frac{\pi ^{d/2}}{\Gamma \left( d/2+1\right) }R^{d}
\label{eq5}
\end{equation}%
the volume of a $d$-dimensional sphere with \textit{radius} $R$. $\Gamma (x)$
is the Gamma function.

MCT provides an equation of motion for the intermediate scattering function $%
S(k,t)$ \cite{GOE09}. For a one-component liquid with Brownian dynamics the
MCT equation for the normalized correlator $\phi (k,t)=S(k,t)/S(k)$ is given
by:%
\begin{equation}
\gamma _{k}\overset{\text{{\Large .}}}{\phi }\left( k,t\right) +\phi \left(
k,t\right) +\int_{0}^{t}dt^{\prime }m\left( k,t-t^{\prime }\right) \overset{%
\text{{\Large .}}}{\phi }\left( k,t^{\prime }\right) =0.  \label{eq6}
\end{equation}%
$\gamma _{k}$ is a microscopic relaxation rate related to the short time
diffusion constant. The memory kernel in bi-polar coordinates reads:%
\begin{eqnarray}
&&m\left( k,t\right) \equiv \mathcal{F}_{k}\left[ \phi \left( q,t\right) %
\right] =\Omega _{d-1}\frac{1}{\left( 4\pi \right) ^{d}}\cdot  \notag \\
&&\cdot \int_{0}^{\infty }dp\int_{\left\vert k-p\right\vert }^{k+p}dqV\left(
k,p,q\right) \phi \left( p,t\right) \phi \left( q,t\right)  \label{eq7}
\end{eqnarray}%
with the vertices in \textit{arbitrary} dimensions $d$ \cite{BAY07}:%
\begin{eqnarray}
&&V\left( k,p,q\right) =n~\frac{pq}{k^{d+2}}S\left( k\right) S\left(
p\right) S\left( q\right) \cdot  \notag \\
&&\cdot \left[ 4k^{2}p^{2}-\left( k^{2}+p^{2}-q^{2}\right) ^{2}\right]
^{\left( d-3\right) /2}\cdot  \label{eq8} \\
&&\cdot \left[ \left( k^{2}+p^{2}-q^{2}\right) c\left( p\right) +\left(
k^{2}-p^{2}+q^{2}\right) c\left( q\right) \right] ^{2}.  \notag
\end{eqnarray}%
$c(k)$ is the direct correlation function and
$\Omega _{d}=2\pi ^{d/2}/\Gamma (d/2)$ is the surface area of a $d$%
-dimensional unit sphere.

The corresponding equation of motion for the self correlator $\phi
^{(s)}(k,t)$ follows from Eq.~(\ref{eq6}) by replacing $\gamma _{k}$ and $%
m(k,t)$ by $\gamma _{k}^{(s)}$ and $m^{(s)}(k,t)$, respectively. $%
m^{(s)}(k,t)$ is given by:%
\begin{eqnarray}
&&m^{\left( s\right) }\left( k,t\right) \equiv \mathcal{F}_{k}^{\left(
s\right) }\left[ \phi \left( q,t\right) ,\phi ^{\left( s\right) }\left(
q,t\right) \right] =\Omega _{d-1}\frac{1}{\left( 4\pi \right) ^{d}}\cdot 
\notag \\
&&\cdot \int_{0}^{\infty }dp\int_{\left\vert k-p\right\vert
}^{k+p}dqV^{\left( s\right) }\left( k,p,q\right) \phi \left( p,t\right) \phi
^{\left( s\right) }\left( q,t\right)  \label{eq9}
\end{eqnarray}%
with the corresponding vertices \cite{BAY07}:%
\begin{eqnarray}
&&V^{\left( s\right) }\left( k,p,q\right) =2n\frac{pq}{k^{d+2}}S\left(
p\right) \left[ \left( k^{2}+p^{2}-q^{2}\right) c\left( p\right) \right]
^{2}\cdot  \notag \\
&&\cdot \left[ 4k^{2}p^{2}-\left( k^{2}+p^{2}-q^{2}\right) ^{2}\right]
^{\left( d-3\right) /2}.  \label{eq10}
\end{eqnarray}%
Note, that the vertices Eq.~(\ref{eq8}) and Eq.~(\ref{eq10}) reduce to the
well-known expressions \cite{GOE09} for $d=3$ for which the triple direct
correlation function $c^{(3)}(k,p,q)$ has been neglected.

\subsection{Static correlation functions}

\begin{figure}[h]
\centerline{%
\includegraphics[width=8cm,clip=true]{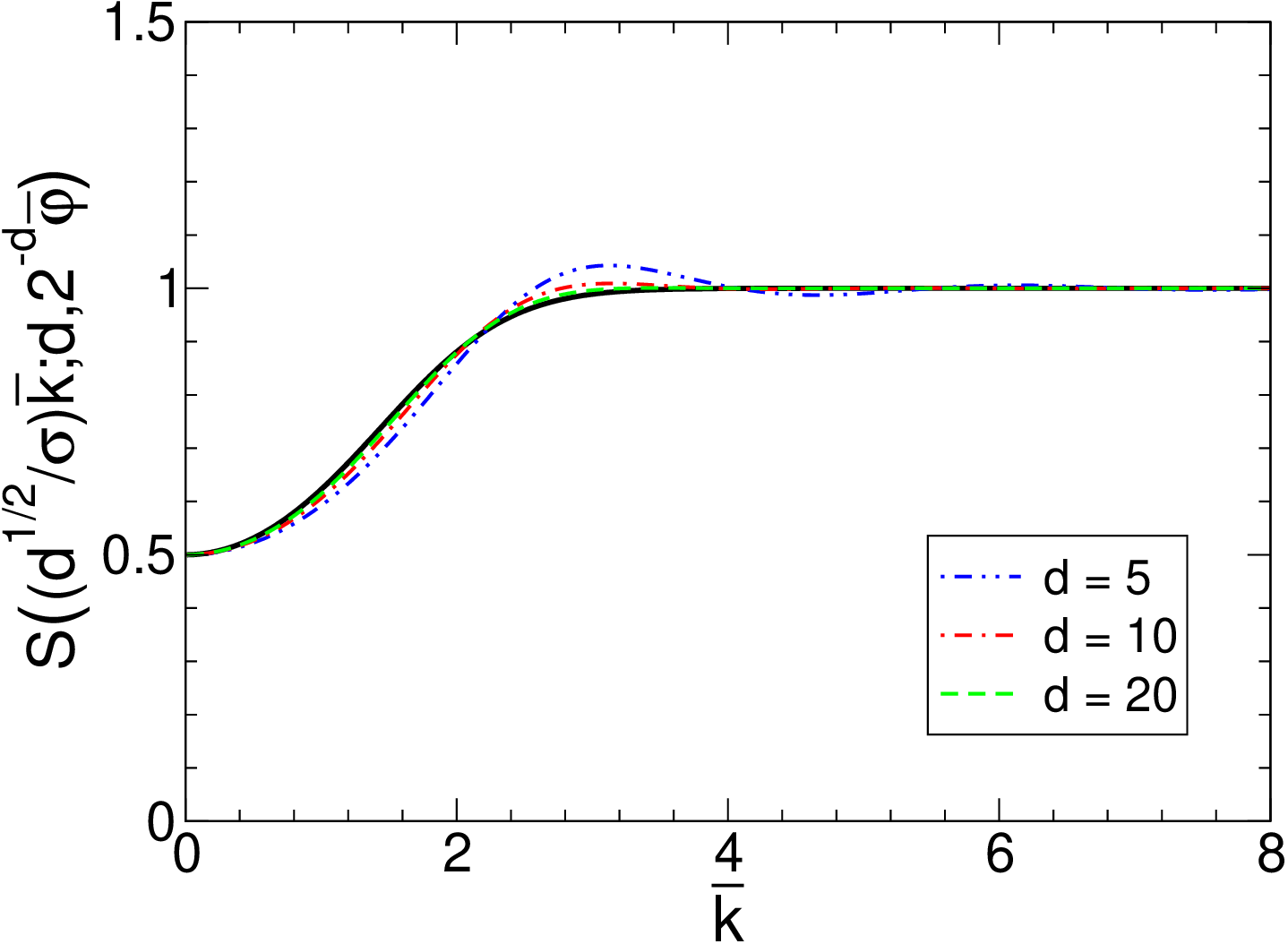}}
\caption{(colour online) $S\left( (\protect\sqrt{d}/\protect\sigma )\bar{k};d,2^{-d}\bar{%
\protect\varphi}\right) $ on the scale $\bar{k}$ for $\bar{\protect\varphi}%
=1 $ and $d=5$, $10$, $20$. The bold black line
is $\bar{S}\left( \bar{k};1\right) $.}
\label{fig1}
\end{figure}

The static correlation function $S(k)\equiv S(k,t=0)$ is related to the
direct correlation function by the Ornstein-Zernike equation:%
\begin{equation}
S\left( k;d,\varphi \right) =\left[ 1-n\left( \varphi \right) c\left(
k;d,\varphi\right) \right] ^{-1}.  \label{eq11}
\end{equation}%
The direct correlation
function $c(k;d,\varphi)$ is known analytically for $d\rightarrow \infty $, in case
that the third and higher virial terms of the virial expansion can be neglected. It has been shown 
\cite{FP99} that the truncation for $d\rightarrow \infty $ at the second
virial term is even valid above the packing fraction at which the virial
series diverges. Under these conditions it is $c(r;d,\varphi )\cong -\Theta (\sigma
-r)=f(r)$ (Mayer function) from which one obtains%
\begin{equation}
c(k;d,\varphi) \cong c\left( k;d\right) = -\left( 2\pi \right) ^{d/2}\sigma ^{d}J_{d/2}\left(
k\sigma \right) /\left( k\sigma \right) ^{d/2}  \label{eq1}
\end{equation}%
where $c(k;d)$ does not depend on $\varphi$. $\sigma $ is the diameter of
the hard spheres and $J_{n}(x)$ the Bessel function of order $n$.
Note, that the $d$- and $\varphi $-dependence of the various quantities is
made explicit in cases where it is useful, and suppressed otherwise. There
are two $d$-dependent $k$-scales on which the $k$-variation of $%
S(k;d,\varphi )$ is different. For%
\begin{equation}
\bar{k}=k\sigma /\sqrt{d}\text{,\hspace{0.5cm}}\bar{\varphi}=2^{d}\varphi
\label{eq12}
\end{equation}%
it follows from Eqs.~(\ref{eq4}),~(\ref{eq11}) and ~(\ref{eq1}) by using the
Taylor series for $J_{d/2}(\sqrt{d}~\bar{k})$ \cite{ABRA70} in the scaling
limit $k\rightarrow \infty $, $d\rightarrow \infty $, $\varphi \rightarrow 0$
such that $\bar{k}=k\sigma /\sqrt{d}$ and $\bar{\varphi}=2^{d}\varphi $ are
fixed:%
\begin{equation}
\lim_{d\rightarrow \infty }S\left( \left( \sqrt{d}/\sigma \right) \bar{k}%
;d,2^{-d}\bar{\varphi}\right) =\left[ 1-\bar{\varphi}~\bar{c}\left( \bar{k}%
\right) \right] ^{-1}\equiv \bar{S}\left( \bar{k};\bar{\varphi}\right)
\label{eq13}
\end{equation}%
where:%
\begin{equation}
\bar{c}\left( \bar{k}\right) =-\exp \left( -\frac{1}{2}\bar{k}^{2}\right) .
\label{eq14}
\end{equation}%
Figure \ref{fig1} demonstrates the convergence of $S\left( k;d,\varphi
\right) $ to the scaling function $\bar{S}(\bar{k};\bar{\varphi})$ on the
scale $\bar{k}$ for $\bar{\varphi}=2^{d}\varphi $ fixed.

$\bar{S}(\bar{k};\bar{\varphi})$ does not exhibit peaks. It increases
monotonically from $\bar{S}(0;\bar{\varphi})=1/[1+\bar{\varphi}]<1$ to the
ideal gas behavior $\bar{S}(\bar{k};\bar{\varphi})=1$ for $\bar{k}%
\rightarrow \infty $. This is the well-known effect for $k\rightarrow 0 
$ of the suppression of the compressibility below the corresponding value of
an ideal gas which, however, is much weaker than for $d=3$.

The second $k$-scale is linear in $d$:%
\begin{equation}
\tilde{k}=k\sigma /d\text{,\hspace{0.5cm}}\tilde{\varphi}=\varphi
2^{d}e^{-d/2}  \label{eq14a}
\end{equation}%
Making use of the asymptotic expansion of $J_{n}(nx)$ \cite{ABRA70} one
obtains:%
\begin{equation}
S\left( (d/\sigma )\tilde{k};d,2^{-d}e^{d/2}\tilde{\varphi}\right) \cong %
\left[ 1-\tilde{\varphi}\tilde{c}_{d}\left( \tilde{k}\right) \right]
^{-1}\equiv \tilde{S}_{d}\left( \tilde{k};\tilde{\varphi}\right)
\label{eq15}
\end{equation}%
where%
\begin{eqnarray}
&&\tilde{c}_{d}\left( \tilde{k}\right) \cong -\frac{1}{\tilde{k}^{d/2}}%
\left( 1-4\tilde{k}^{2}\right) ^{-1/4}\cdot \\
&&\cdot \exp \left[ -\frac{d}{2}\left( \mathrm{arctanh}\sqrt{1-4\tilde{k}^{2}%
}-\sqrt{1-4\tilde{k}^{2}}\right) \right]  \notag
\end{eqnarray}%
for $\tilde{k}<1/2$, and%
\begin{eqnarray}
&&\tilde{c}_{d}\left( \tilde{k}\right) \cong -\frac{2}{\tilde{k}^{d/2}}%
\left( 4\tilde{k}^{2}-1\right) ^{-1/4}\cdot \\
&&\cdot \cos \left[ \frac{d}{2}\sqrt{4\tilde{k}^{2}-1}-\frac{d}{2}\arctan 
\sqrt{4\tilde{k}^{2}-1}-\frac{\pi }{4}\right]  \notag
\end{eqnarray}%
for $\tilde{k}>1/2$. $\tilde{S}_{d}(\tilde{k};\tilde{\varphi})$ oscillates
for $\tilde{k}>1/2$.

\begin{figure}[h]
\centerline{%
\includegraphics[width=8cm,clip=true]{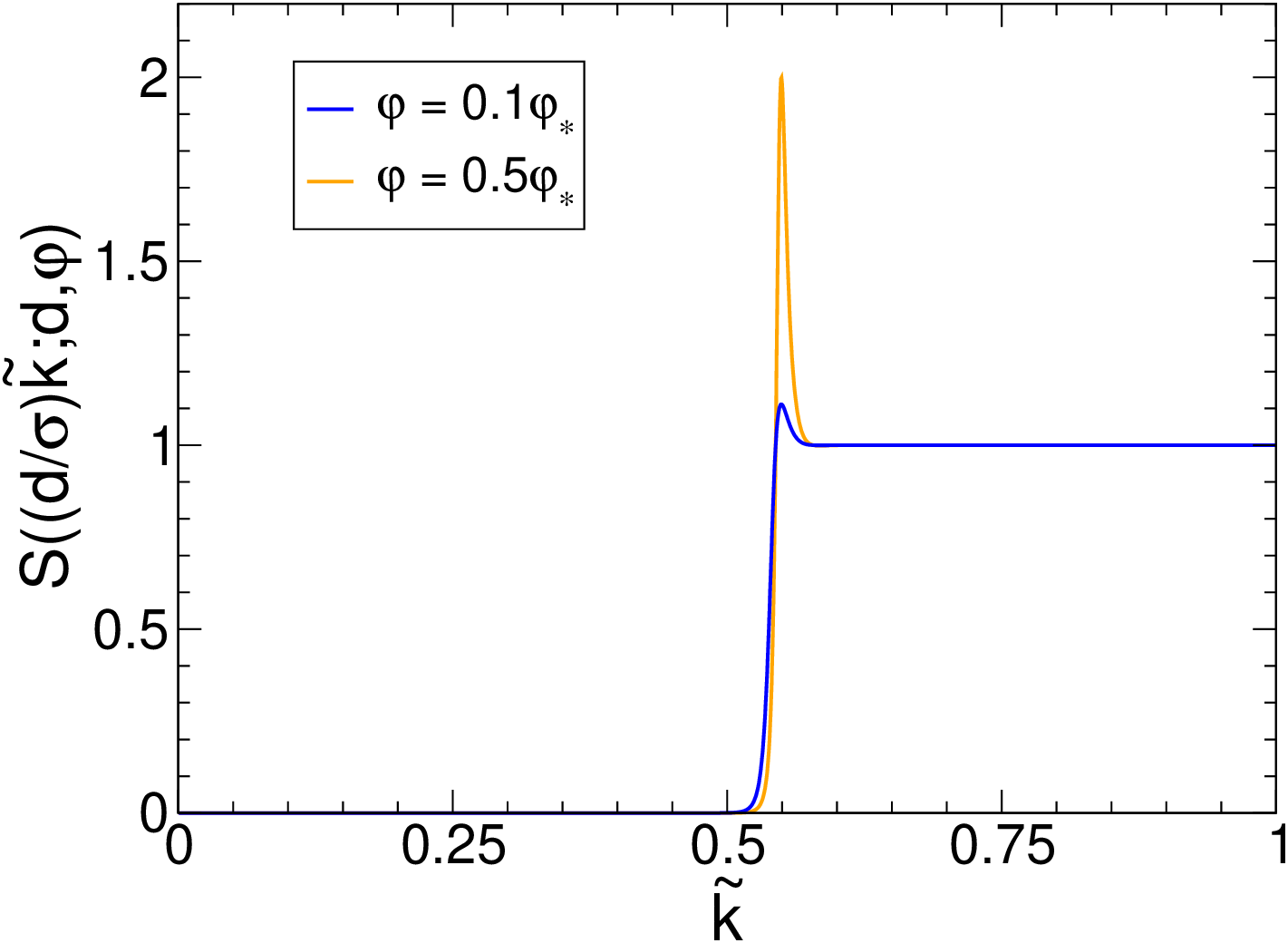}}
\caption{(colour online) $S\left( (d/\protect\sigma )\tilde{k};d,\protect%
\varphi \right) $ versus $\tilde{k}$ for $d=200$ and $\protect\varphi =0.1%
\protect\varphi _{\ast }$, $\protect\varphi =0.5\protect\varphi %
_{\ast }$.}
\label{fig2}
\end{figure}

The position $k_{\ast }(d)$ of the main peak (first sharp diffraction peak)
of $S(k;d,\varphi )$ is given by the first nonvanishing zero of $%
J_{d/2+1}(x) $, which is \cite{ABRA70}:%
\begin{equation}
k_{\ast }\left( d\right) \cong \left( d/2+1\right) +a_{0}\left( d/2+1\right)
^{1/3}\text{, \ \ }a_{0}=1.8557571.  \label{eq17}
\end{equation}%
Since $k_{\ast }(d)$ and $c(k;d)$ from Eq. (\ref{eq1}) are $\varphi $-independent the packing
fraction $\varphi _{\ast }(d)$ for which $S(k_{\ast }(d);d,\varphi )$
diverges follows from Eqs.~(\ref{eq4}),~(\ref{eq5}) and ~(\ref{eq11}):%
\begin{equation}
\varphi _{\ast }\left( d\right) =\frac{\pi ^{d/2}\left( \sigma /2\right) ^{d}%
}{\Gamma \left( d/2+1\right) c\left( k_{\ast }\left( d\right) ;d\right) }.
\label{eq18}
\end{equation}%
Using again the asymptotic properties of the Gamma and Bessel function and
especially from \cite{ABRA70} 
\begin{eqnarray}
J_{d/2}\left( k_{\ast }\left( d\right) \right) &=&J_{d/2+1}^{\prime }\left(
k_{\ast }\left( d\right) \right) \cong -b_{0}\cdot \left( d/2+1\right)
^{-2/3},  \notag \\
b_{0} &\cong &1.1131028
\end{eqnarray}%
we obtain:%
\begin{eqnarray}
\varphi _{\ast }\left( d\right) &\cong &c_{0}\cdot d^{1/6}\exp \left(
a_{0}\left( d/2\right) ^{1/3}\right) \left( \sqrt{8/e}\right) ^{-d},  \notag
\\
c_{0} &=&b_{0}^{-1}\pi ^{-1/2}2^{-2/3}e\cong 0.867956,  \label{eq19}
\end{eqnarray}%
i.e. the leading $d$-dependence of $\varphi _{\ast }\left( d\right) $ is the
exponential factor $(\sqrt{8/e})^{-d}\cong (1.7155)^{-d}$ \cite{footnote1}.
Note that $S\left( k;d,\varphi \right) >0$ for all $k$ provided $\varphi
<\varphi _{\ast }\left( d\right) $. For $\varphi \ll \varphi _{\ast }\left(
d\right) $ the height of the first sharp diffraction peak of $S(k;d)$ is
given as%
\begin{equation}
S\left( k_{\ast }(d);d,\varphi \right) \cong 1+\frac{\varphi }{\varphi
_{\ast }\left( d\right) }.
\end{equation}%
Figure~\ref{fig2} presents $S(k;d,\varphi )$ on the scale $\tilde{k}=k\sigma
/d$ for $d=200$ and $\varphi $ of order $\varphi _{\ast }(d)$.

\begin{figure}[h]
\centerline{%
\includegraphics[width=8cm,clip=true]{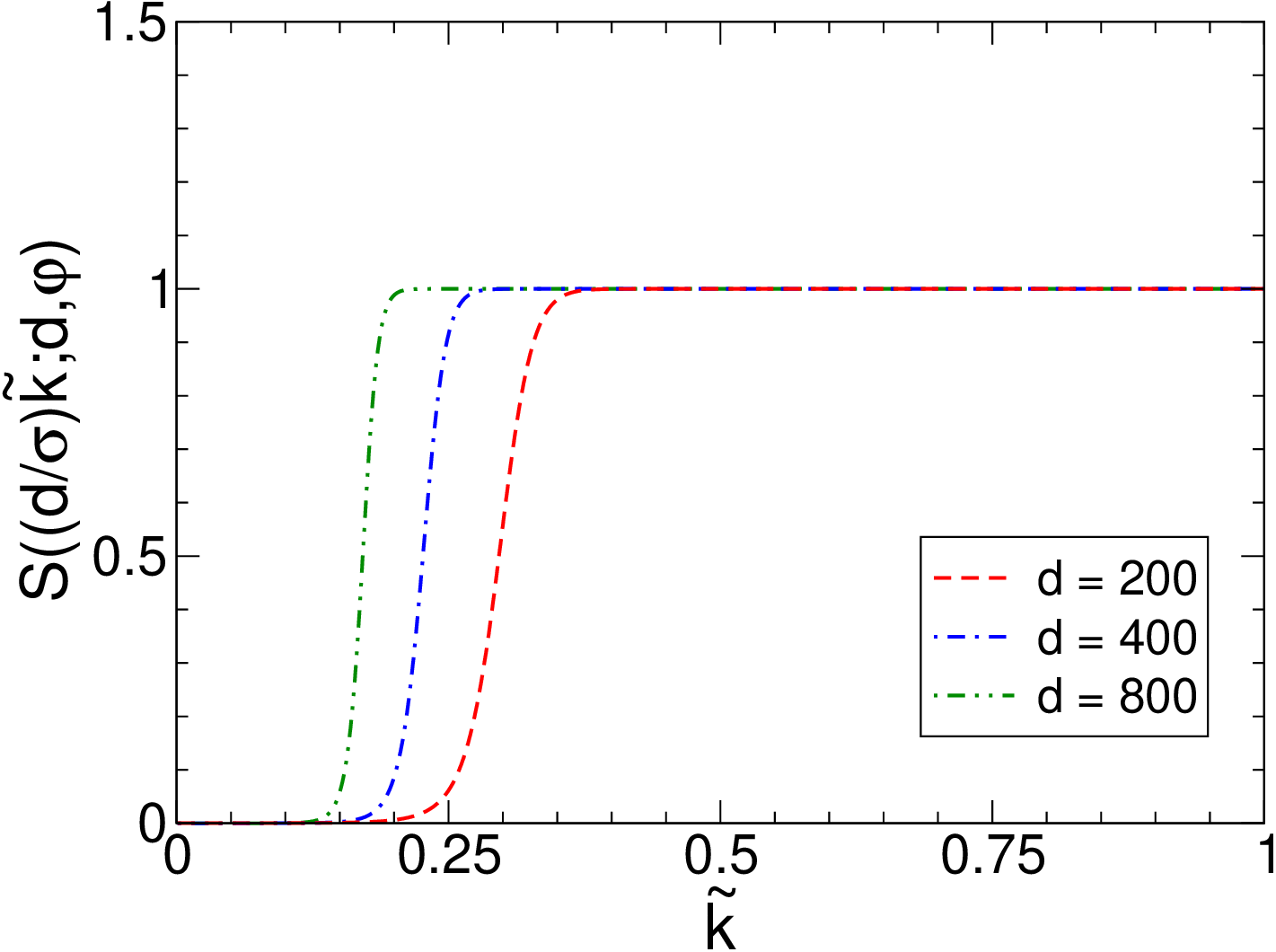}}
\caption{(colour online) $S\left( (d/\protect\sigma )\tilde{k};d,\protect%
\varphi \right) $ versus $\tilde{k}$ for $d=200$, $400$, $800$
and $\protect\varphi =\protect\varphi _{c}\left( d\right) $.}
\label{fig3}
\end{figure}

$S((d/\sigma )\tilde{k};d,\varphi )$ is practically zero for $\tilde{k}%
\lesssim 0.5$, develops its main peak at $\tilde{k}_{\ast }(d)=k_{\ast
}(d)\sigma /d$ and decays very fast to one for $\tilde{k}>\tilde{k}_{\ast
}(d)$. In contrast to this, Figure \ref{fig3} shows $S((d/\sigma )\tilde{k}%
;d,\varphi )$ again on the scale $\tilde{k}$ but for $\varphi $ of order $%
\varphi _{c}(d)\cong 0.22\cdot d^{2}2^{-d}$, which will turn out to be the
critical packing fraction of the MCT glass transition. Note that
$\varphi _{c}\left( d\right) $ is exponentially smaller than $%
\varphi _{\ast }\left( d\right) $.

Except for $\tilde{k}=O(1/\sqrt{d})$ which is $k\sigma =O(\sqrt{d})$ it is $%
S(k;d,\varphi )\cong 1$, i.e. the static structure factor on a $k$-scale 
\textit{linear} in $d$ is very close to that of an ideal gas for $d\rightarrow \infty $%
. The first sharp diffraction peak of the conventional liquids has
disappeared due to the lack of intermediate range order at $\varphi =\varphi
_{c}(d)$. These results will be used for the analytical treatment of the
MCT equation. It is important to note that $[S(k;d,\varphi)-1]$ is very small
for $k\sigma=O(d)$, $\varphi=O(\varphi_c(d))$ and $d\gg 1$, but not zero.
Accordingly the direct correlation function does not vanish, in contrast to
an ideal gas with finite density. Therefore the vertices Eq.~(\ref{eq8}) and
Eq.~(\ref{eq10}) are nonzero and exhibit nontrivial $k$-dependence.

\subsection{Numerical solution}

The nonergodicity parameter for the collective correlator is the long time limit of the
normalized intermediate scattering function%
\begin{equation}
f\left( k;d,\varphi \right) =\lim_{t\rightarrow \infty }\phi \left(
k,t;d,\varphi \right)  \label{eq20}
\end{equation}%
and similarly for the self correlator%
\begin{equation}
f^{\left( s\right) }\left( k;d,\varphi \right) =\lim_{t\rightarrow \infty
}\phi ^{\left( s\right) }\left( k,t;d,\varphi \right) .  \label{eq21}
\end{equation}%
They are the order parameters for the liquid-glass transition.
From Eq.~(\ref{eq6}) and the corresponding equation for $\phi
^{(s)}(k,t;d,\varphi )$ one obtains the algebraic, nonlinear equations for
the nonergodicity parameters%
\begin{equation}
f\left( k;d,\varphi \right) /\left[ 1-f\left( k;d,\varphi \right) \right] =%
\mathcal{F}_{k}\left[ f\left( q;d,\varphi \right) ;d,\varphi \right]
\label{eq22}
\end{equation}%
and a similar equation for $f^{(s)}(k;d,\varphi )$ by replacing $\mathcal{F}%
_{k}$ by $\mathcal{F}_{k}^{(s)}$. Note, that we also made explicit the $d$-
and $\varphi $-dependence of the functional $\mathcal{F}_{k}$ on the r.h.s.
of Eq.~(\ref{eq22}). Eq.~(\ref{eq22}) and the corresponding one for $%
f^{(s)}(k;d,\varphi )$ has been solved numerically as follows.

Eq. (\ref{eq22}) is rewritten such that the nonergodicity parameters $%
f\left( k;d,\varphi \right) $ can be evaluated by iterating the equation%
\begin{equation}
f^{\left( i+1\right) }\left( k;d,\varphi \right) =\frac{\mathcal{F}_{k}\left[
f^{\left( i\right) }\left( q;d,\varphi \right) ;d,\varphi \right] }{\mathcal{%
F}_{k}\left[ f^{\left( i\right) }\left( q;d,\varphi \right) ;d,\varphi %
\right] +1}  \label{a}
\end{equation}%
with the initial value%
\begin{equation}
f^{\left( 0\right) }\left( k;d,\varphi \right) \equiv 1  \label{b}
\end{equation}%
and similar equations for $f^{\left( s\right) }\left( k;d,\varphi \right) $.
Note, in case of hard spheres the functional $\mathcal{F}_{k}$ for the zero
order iterate $f^{(0)}\left( k;d,\varphi \right) $ from Eq. (\ref{b}) exists
only for a finite cut-off at $k_{\max }$. The integrals appearing in $%
\mathcal{F}_{k}\left[ f^{\left( i\right) }\left( k;d,\varphi \right)
;d,\varphi \right] $ are replaced by Riemann sums with an upper cutoff $%
\sigma k_{\max }=\max (40d^{1/2};4d;0.2d^{3/2})$ and 500 gridpoints for $%
d<200$, 1000 gridpoints for $200\leq d\leq 600$ and 1500 gridpoints for $%
d>600$. The critical packing fraction $\varphi _{c}(d)$ is the packing
fraction, where%
\begin{equation}
f\left( k;d,\varphi \right) \left\{ 
\begin{array}{cc}
=0 & ,~\varphi <\varphi _{c}(d) \\ 
\neq 0 & ,~\varphi \geq \varphi _{c}(d)%
\end{array}%
\right.
\end{equation}%
and the critical nonergodicity parameters are given by%
\begin{equation}
f_{c}\left( k;d\right) \equiv f(k;d,\varphi _{c}(d)).
\end{equation}%
Because the real critical packing fraction and the critical nonergodicity parameters can never be
computed numerically in finite time, we evaluated $f_{c}\left( k;d\right) $ at a
packing fraction $\hat{\varphi}_{c}(d)$ where $\lim_{i\rightarrow \infty
}f^{\left( i\right) }\left( k;d,\hat{\varphi}_{c}(d)\right) =0$ but 
\begin{equation}
\min_{i}\left( \max_{k}\left\vert \frac{f^{\left( i+1\right) }\left( k;d,%
\hat{\varphi}_{c}(d)\right) -f^{\left( i\right) }\left( k;d,\hat{\varphi}%
_{c}(d)\right) }{f^{\left( i+1\right) }\left( k;d,\hat{\varphi}%
_{c}(d)\right) }\right\vert \right) <\varepsilon
\end{equation}%
with $\varepsilon =10^{-7}$ for $d\leq 600$ and $\varepsilon =10^{-5}$ for $%
d>600$. It can be estimated that the relative difference between this
packing fraction $\hat{\varphi}_{c}(d)$ and the real critical packing
fraction $\varphi _{c}(d)$ is of order $\varepsilon $. It has been verified
that the system really becomes nonergodic near this packing fraction, i.e. $%
f\left( k;d,\varphi \right) \neq 0$ for $\varphi >(1+\varepsilon \cdot O(1))%
\hat{\varphi}_{c}(d)$. The critical nonergodicity parameters can then be
approximated by $f_{c}\left( k;d\right) \cong f^{\left( i_{0}\right) }\left(
k;d,\hat{\varphi}_{c}(d)\right) $ where $i_{0}$ equals the iteration step,
where 
\begin{equation}
\max_{k}\left\vert \frac{f^{\left( i_{0}+1\right) }\left( k;d,\hat{\varphi}%
_{c}(d)\right) -f^{\left( i_{0}\right) }\left( k;d,\hat{\varphi}%
_{c}(d)\right) }{f^{\left( i_{0}+1\right) }\left( k;d,\hat{\varphi}%
_{c}(d)\right) }\right\vert
\end{equation}%
reaches a minimum \cite{WINK00}. It has been verified that there are no
visible differences in the critical nonergodicity parameters obtained by
this procedure with different values of $\varepsilon $ and that $%
f(k;d,\varphi _{c}\left( d\right) +\Delta \varphi )$ converges to $f^{\left(
i_{0}\right) }\left( k;d,\hat{\varphi}_{c}(d)\right) $ with order $\sqrt{%
\Delta \varphi }$. Additionally it has been verified that $f_{c}\left(
k_{\max };d\right) <10^{-16}$ for all evaluated dimensions. By increasing $%
k_{\max }$ and the number of gridpoints the relative error of the critical
packing fraction due to the discretization can be estimated to be smaller
than $10^{-3}$ for $d\leq 600$.

The nonergodicity parameters always show numerical artifacts on the first
few gridpoints in $k$-space. This is a problem when trying to observe the
characteristics of $f_{c}\left( k;d\right) $ for small wavenumbers,
especially for high dimensions. So we interpolated $f_{c}\left( k;d\right) $
onto a much finer $k$-grid and performed one single iteration step
equivalent to the one given in Eq. (\ref{a}). This procedure improves the
result for $f_{c}\left( k;d\right) $ by shifting the numerical artifacts to
much smaller values of $k$.

\begin{figure}[h]
\centerline{%
\includegraphics[width=8.5cm,clip=true]{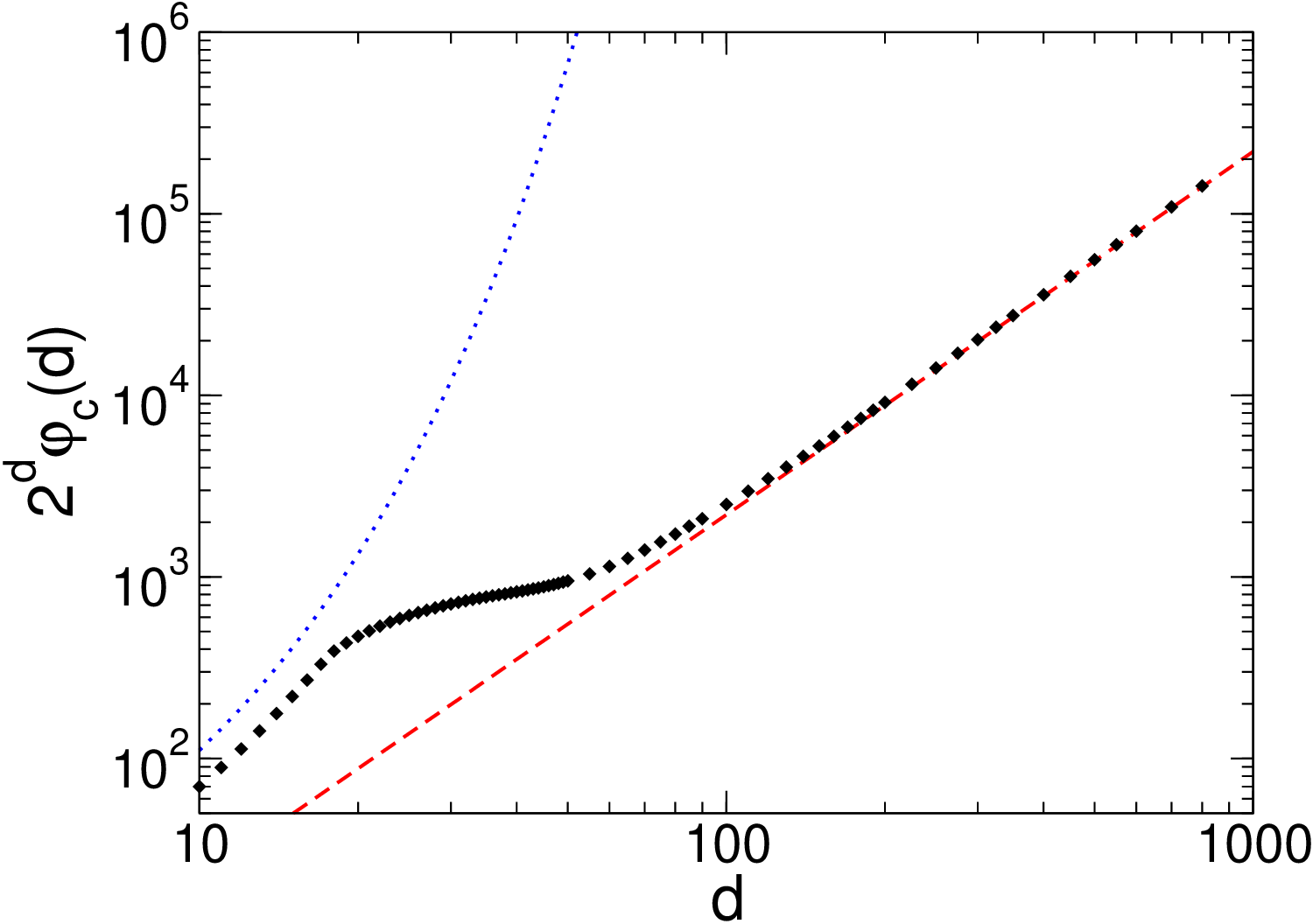}}
\caption{(colour online) $d$-dependence of the critical packing fraction $\protect\varphi %
_{c}\left( d\right) $ on a log-log representation. The dashed line is $2^{d}%
\protect\varphi _{c}(d)\protect\cong ad^{2}$. The dotted line is $\protect%
\varphi _{\ast }\left( d\right) $ from Eq. (\protect\ref{eq18}).}
\label{fig4}
\end{figure}

From this solution we obtain the critical packing fraction $\varphi _{c}(d)$%
, shown in Figure~\ref{fig4}.

The $d$-dependence of $\varphi _{c}$ can be well fitted by%
\begin{equation}
\varphi _{c}\left( d\right) \cong ad^{2}2^{-d}\text{,\hspace{0.5cm}}a\cong
0.22.  \label{eq23}
\end{equation}%

\begin{figure}[h]
\centerline{%
\includegraphics[width=8.5cm,clip=true]{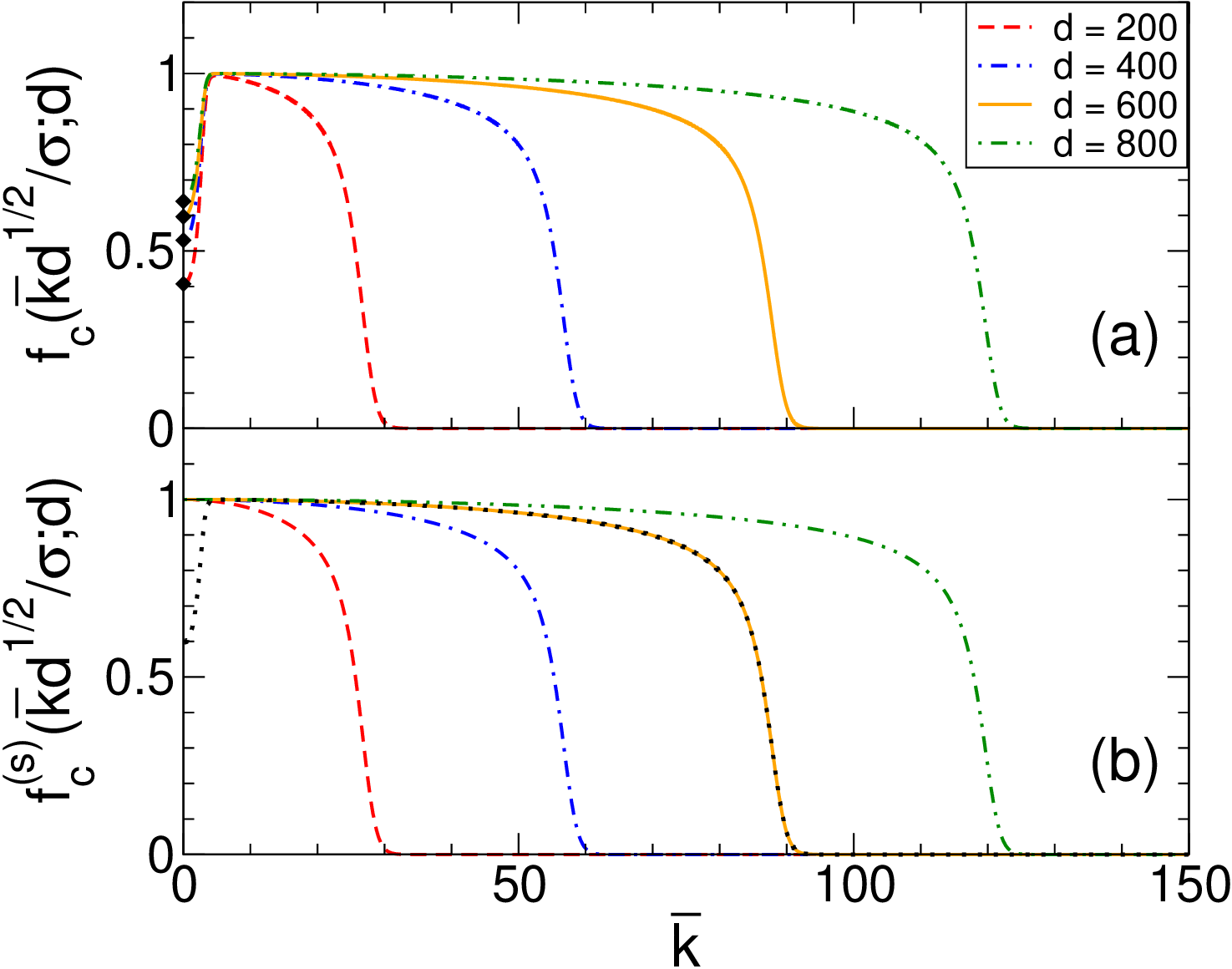}}
\caption{(colour online) $\bar{k}$-dependence of the critical nonergodicity
parameters $f_{c}\left( k;d\right) $ (a) and $f_{c}^{\left( s\right) }\left(
k;d\right) $ (b) for $d=200$, $400$, $600$ and $800$.
Diamonds in (a) mark the numerical precise values for $f_{c}\left(
0;d\right) $ and the dotted line in (b) presents $f_{c}\left( k;600\right) $.}
\label{fig5}
\end{figure}

The critical nonergodicity parameters $f_{c}(k;d)$ and $f_{c}^{(s)}(k;d)$
are presented in Figure~\ref{fig5}a and ~\ref{fig5}b, respectively, for
different values of $d$.

Figure \ref{fig5}a and ~\ref{fig5}b reveal the following properties:

\begin{itemize}
\item[(i)] $f_{c}(k;d)$ and $f_{c}^{(s)}(k;d)$ differ on the scale $k\sigma
=O(\sqrt{d})$, but become identical on the scale $k\sigma =O(d)$, for $d$
large enough.

\item[(ii)] Both, $f_{c}(k;d)$ and $f_{c}^{(s)}(k;d)$, exhibit \textit{%
non-Gaussian} $k$-dependence.

\item[(iii)] There are three characteristic $k$-scales. $f_{c}(k;d)$
increases from $f_{c}(0;d)$ to its maximum value on a scale $k\sigma \sim 
\sqrt{d}$, develops a plateau on a scale $k\sigma \sim d$ and it varies on a
scale $k\sigma \sim d^{3/2}$, on which a steep descent to zero occurs for $%
k\sigma \cong \hat{k}_{0}d^{3/2}$ where $\hat{k}_{0}\cong 0.15$. The plateau
value on scale $k\sigma \sim d$ converges to one, for $d\rightarrow \infty $.

\item[(iv)] Since $f_{c}\left( k;d\right) $ changes from one to zero around $%
k\sigma \cong \hat{k}_{0}d^{3/2}$ we will choose $\hat{k}_{0}$ such that $%
f_{c}(\hat{k}_{0}d^{3/2}/\sigma ;d)=1/2$. $f_{c}\left( k;d\right) $ is of
order one for ($k\sigma -\hat{k}_{0}d^{3/2})$ of order $d^{1/2}$, i.e. for $(%
\tilde{k}-\hat{k}_{0}d^{1/2})$ of order $d^{-1/2}$.
\end{itemize}

Using Eq. (\ref{eq22}), $f_{c}(0;d)$ can be represented as a functional of $%
f_{c}(k;d)$ \cite{BAY07}. Making use of this relationship yields the
numerical precise values for $f_{c}(0;d)$ shown by diamonds in Figure~\ref%
{fig5}a. Note that for the self correlators it is $f_{c}^{(s)}(0;d)=1$ for
all $d$, because the momentum of a tagged particle is not conserved.

\begin{figure}[h]
\centerline{%
\includegraphics[width=8.5cm,clip=true]{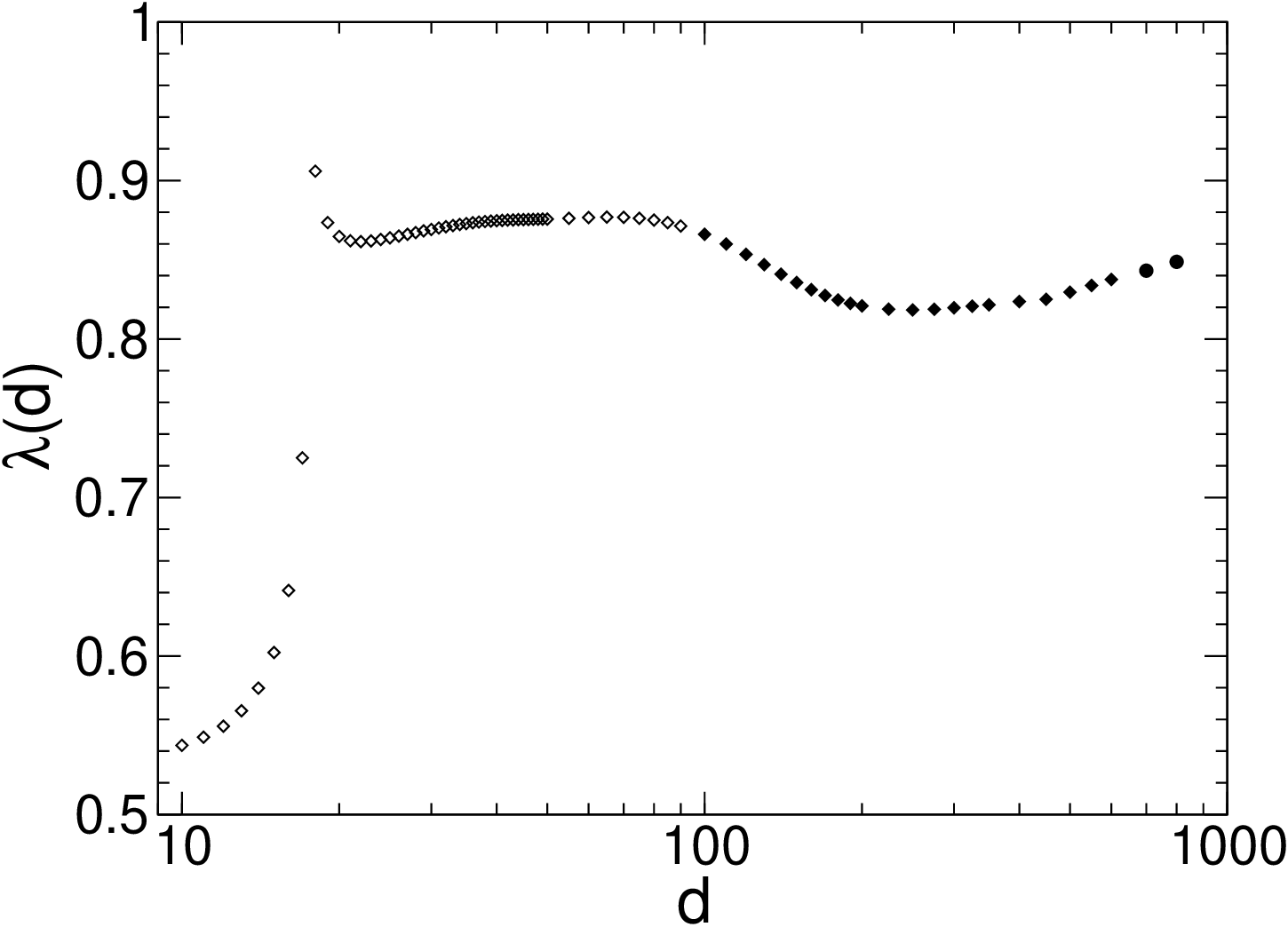}}
\caption{$d$-dependence of the exponent parameter $\protect\lambda $. The full
symbols mark the regime where $\protect\varphi _{c}(d)$ from Figure \protect
\ref{fig4} follows the asymptotic result (\protect\ref{eq23}). The values
for $d=700$ and $800$ depicted by full circles possibly have a larger relative
error.}
\label{fig6}
\end{figure}

A crucial quantity of MCT is the exponent parameter $\lambda (d)$ which
determines the critical exponents of both power laws in time, the critical
law and the von Schweidler law, and the divergence of the corresponding
relaxation time scales at the glass transition singularity \cite{GOE09}.
Figure \ref{fig6} presents the numerical result for $\lambda (d)$ with an
estimated relative error of about $5\cdot 10^{-3}$ for $d\leq 600$ and
possibly a larger error for $d=700$ and $800$.

Since the direct correlation function (\ref{eq1}) is not correct for small $%
d $, the variation of $\lambda $ with $d$ below $100$ and particularly the
high sensitivity close to $d=17$ is an artifact of the incorrect static
input. This holds also for $\varphi _{c}(d)$ of Figure \ref{fig4}. The concave
curvature of $\varphi _{c}\left( d\right) $ and the cusp-like behavior of $%
\lambda \left( d\right) $ around $d=17$ relates to the fact that the glass
transition is still influenced by the main diffraction peak of the static
structure factor for $d\lesssim 17$, while this peak is not important any
more for $d\gtrsim 17$. Figure \ref{fig4} reveals the correct asymptotic $d$%
-dependence to appear for $d\gtrsim 100$.

\section{MCT EQUATIONS: ANALYTICAL APPROACH}

In this section we will demonstrate that the MCT functional $\mathcal{F}_{k}$
strongly simplifies for $d\rightarrow \infty $. The essential steps will be
given, only.

In a first step we rewrite $\mathcal{F}_{k}[f(q)]$ (Eq.~(\ref{eq7}) with $%
\phi (q,t)$ replaced by the nonergodicity parameters $f(q)$) on the scale $\tilde{k}=k\sigma /d$.
Quantities on this scale will be denoted by a tilde, e.g.~$\tilde{f}(\tilde{k%
})$. Now we prove that $\mathcal{F}[f(q)]$ on this scale reduces to $%
\mathcal{F}_{k}^{(s)}$ for $\tilde{k}$ of order one and larger.

The last square bracket in Eq.~(\ref{eq8}) contains a mixed term $\sim c(%
\tilde{p}d/\sigma)c(\tilde{q}d/\sigma)$ which is oscillating in $\tilde{p}$ and $\tilde{q}$
faster and faster under an increase of $d$. Since $S(\tilde{p}d/\sigma)$ $S(\tilde{q%
}d/\sigma)$ and $f(\tilde{p}d/\sigma)f(\tilde{q}d/\sigma)$ are smooth and not strongly
oscillating (see section II) the mixed term integrated over $\tilde{p}$ and $%
\tilde{q}$ will not contribute for $d\rightarrow \infty $. Taking account of
this fact and using the integrand's symmetry with respect to $\tilde{p}%
\leftrightarrow \tilde{q}$ we get for $d\rightarrow \infty $ the
MCT functional Eq.~(\ref{eq7}) but with the replacement:%
\begin{eqnarray}
&&V\left( \tilde{k}d/\sigma ,\tilde{p}d/\sigma ,\tilde{q}d/\sigma \right)
\rightarrow  \notag \\
&\rightarrow &S\left( \tilde{k}d/\sigma \right) S\left( \tilde{q}d/\sigma
\right) V^{\left( s\right) }\left( \tilde{k}d/\sigma ,\tilde{p}d/\sigma ,%
\tilde{q}d/\sigma \right)  \notag \\
&\rightarrow &V^{\left( s\right) }\left( \tilde{k}d/\sigma ,\tilde{p}%
d/\sigma ,\tilde{q}d/\sigma \right) .  \label{eq24}
\end{eqnarray}%
The latter step in Eq. (\ref{eq24}) uses $S(\tilde{k}d;d)\rightarrow 1$ for $%
d\rightarrow \infty $ and $\tilde{k}$ of order one or larger. Remember that
this does not hold for $\tilde{k}=O(1/\sqrt{d})$ (see section II. B). Eq. (%
\ref{eq24}), together with the Vineyard approximation for $\phi \left( 
\tilde{p}d/\sigma ,t\right) $ in $\mathcal{F}_{k}^{\left( s\right) }$ (Eq. (%
\ref{eq9})), implies that the MCT equations for the self and collective
correlator on a $k$-scale linear in $d$ become identical for $d\rightarrow
\infty $. However, for large but finite $d$, there is an interval $k\sigma
\in \lbrack 0,O(\sqrt{d})]$ for which both MCT functionals differ from each
other.

Having reduced $V$ to $V^{\left( s\right) }$ on the $k$-scale linear in $d$
further simplifications occur due to $d\rightarrow \infty $. First of all we
can replace $S\left( \tilde{p}d/\sigma \right) $ in $V^{\left( s\right) }(%
\tilde{k}d/\sigma ,\tilde{p}d/\sigma ,\tilde{q}d/\sigma )$ by one. The
product of both square brackets in (\ref{eq10}) can be rewritten as%
\begin{eqnarray}
&&\sigma ^{-2d+2}d^{2d-2}\left( 2\tilde{k}\tilde{p}\right) ^{d-1}\left[
1-\left( \frac{\tilde{k}^{2}+\tilde{p}^{2}-\tilde{q}^{2}}{2\tilde{k}\tilde{p}%
}\right) ^{2}\right] ^{\frac{d-3}{2}}\cdot  \notag \\
&&\cdot \left( \frac{\tilde{k}^{2}+\tilde{p}^{2}-\tilde{q}^{2}}{2\tilde{k}%
\tilde{p}}\right) ^{2}c^{2}\left( \tilde{p}d/\sigma \right) .  \label{eq25}
\end{eqnarray}%
The square bracket in (\ref{eq25}) can be replaced by $e^{-\frac{d}{2}\left( 
\frac{\tilde{k}^{2}+\tilde{p}^{2}-\tilde{q}^{2}}{2\tilde{k}\tilde{p}}\right)
^{2}}$. Then we use%
\begin{equation}
x^{2}e^{-\frac{d}{2}x^{2}}\rightarrow \frac{\sqrt{\pi }}{4}\left( \frac{2}{d}%
\right) ^{3/2}\left[ \delta \left( x-\sqrt{\frac{2}{d}}\right) +\delta
\left( x+\sqrt{\frac{2}{d}}\right) \right]  \label{eq26}
\end{equation}%
which allows to perform the integration over $\tilde{q}$. Next we account
for the asymptotic behavior of $J_{n}\left( nx\right) $ for $n\rightarrow
\infty $ \cite{ABRA70} and obtain from Eq. (\ref{eq1}):%
\begin{eqnarray}
&&c^{2}\left( \tilde{p}d\right) =\left( 2\pi \right) ^{d}\sigma
^{2d}d^{-d}J_{d/2}^{2}\left[ \left( d/2\right) 2\tilde{p}\right] /\tilde{p}%
^{d}  \notag \\
&\cong &4\left( 2\pi \right) ^{d}\sigma ^{2d}d^{-d}\left( \tilde{p}^{d}\pi d%
\sqrt{4\tilde{p}^{2}-1}\right) ^{-1}\Theta \left( \tilde{p}-\frac{1}{2}%
\right) \cdot  \notag \\
&&\cdot \cos ^{2}\left[ \frac{d}{2}\sqrt{4\tilde{p}^{2}-1}-\frac{d}{2}%
\arctan \sqrt{4\tilde{p}^{2}-1}-\frac{\pi }{4}\right] .  \label{eq27}
\end{eqnarray}%
With these simplifications and the fact that $\cos ^{2}\left[ \ldots \right] 
$ in (\ref{eq27}) oscillates very fast for $d$ large with average $1/2$ we
arrive at:%
\begin{equation}
\mathcal{\tilde{F}}_{\tilde{k}}\left[ \tilde{f}\left( \tilde{q}\right) %
\right] \cong \varphi \frac{2^{d}}{\tilde{k}^{2}\pi d}\int_{\frac{1}{2}%
}^{\infty }d\tilde{p}\frac{\tilde{p}}{\sqrt{4\tilde{p}^{2}-1}}\tilde{f}(%
\tilde{p})\left( \tilde{f}(\tilde{q}_{-})+\tilde{f}(\tilde{q}_{+})\right)
\label{eq28}
\end{equation}%
where:%
\begin{equation}
\tilde{q}_{\pm }=\left[ \tilde{k}^{2}+\tilde{p}^{2}\pm 2\sqrt{\frac{2}{d}}%
\tilde{k}\tilde{p}\right] ^{1/2}.  \label{eq29}
\end{equation}

\begin{figure}[h]
\centerline{%
\includegraphics[width=8.5cm,clip=true]{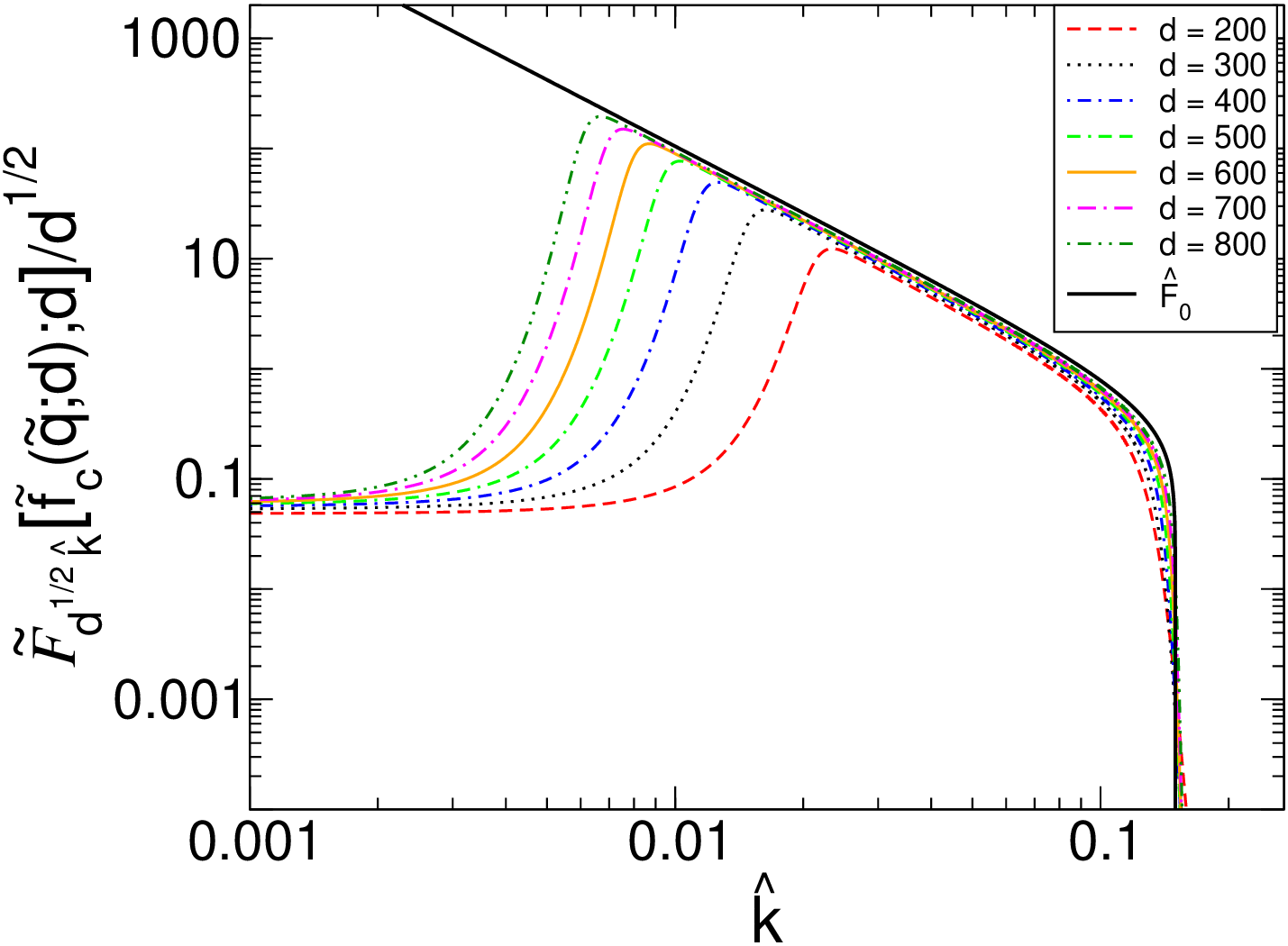}}
\caption{(colour online) The numerical result for $\mathcal{\tilde{F}}_{%
\protect\sqrt{d}\,\hat{k}}\left[ \tilde{f}_{c}\left( \tilde{q};d\right) ;d%
\right] /\protect\sqrt{d}$ as function of $\hat{k}$ for various $d$. The
master function $\hat{F}_{0}(\hat{k})$ is shown as bold black line.}
\label{fig7}
\end{figure}

Our first goal will be the evaluation of the critical packing fraction $%
\varphi _{c}\left( d\right) $. For this we choose $\tilde{k}=\tilde{k}%
_{0}\equiv \hat{k}_{0}d^{1/2}$ such that $\tilde{f}_{c}(\tilde{k}_{0})=1/2$
(see section II. C). Eq. (\ref{eq22}) implies%
\begin{equation}
\mathcal{\tilde{F}}_{\tilde{k}_{0}}\left[ \tilde{f}_{c}\left( \tilde{q}%
\right) \right] =1\text{,\hspace{0.5cm}}\varphi =\varphi _{c}\left( d\right)
.  \label{eq30}
\end{equation}%
$\tilde{f}_{c}(\tilde{q}_{\pm })$ is of order one for $\tilde{k}=\tilde{k}%
_{0} $ and $\tilde{p}=O\left( 1\right) $ and decays rapidly to zero for $%
\tilde{p}\gg 1$. Furthermore $\tilde{f}\left( \tilde{p}\right) \cong 1$ for $%
\tilde{p}=O\left( 1\right) $. Therefore the integral in Eq. (\ref{eq28}) is
of order one, i.e. order $d^{0}$. Then we get from Eq. (\ref{eq28}) with $%
\varphi =\varphi _{c}\left( d\right) $ and $\tilde{k}=\hat{k}\sqrt{d}$:%
\begin{equation}
1\cong \mathcal{\tilde{F}}_{\tilde{k}_{0}}\left[ \tilde{f}_{c}\left( \tilde{q%
}\right) \right] \cong \pi ^{-1}\varphi _{c}\left( d\right) \left(
2^{d}/d^{2}\right) \hat{k}_{0}^{-2}O\left( d^{0}\right) .
\end{equation}%
$\hat{k}_{0}$ is of order $d^{0}$, as well. Consequently it must be:%
\begin{equation}
\varphi _{c}\left( d\right) \cong const\cdot d^{2}2^{-d}
\end{equation}%
in agreement with the numerical result for $d\gtrsim 100$ (cf. Figure \ref%
{fig4}). Next we choose $\tilde{k}\leq \tilde{k}_{0}$. $\tilde{f}_{c}(\tilde{q%
}_{\pm })\approx $ \thinspace $0$ for $\tilde{q}_{\pm }>\tilde{k}_{0}$,
which implies 
\begin{equation*}
\tilde{p}<\mp \sqrt{2/d}~\tilde{k}+\sqrt{\left( 2/d\right) \tilde{k}%
^{2}+\left( \tilde{k}_{0}^{2}-\tilde{k}^{2}\right) }\cong \sqrt{\left( 
\tilde{k}_{0}^{2}-\tilde{k}^{2}\right) }
\end{equation*}%
for $d\rightarrow \infty $ and $\tilde{k}\leq \tilde{k}_{0}$. With $\tilde{f}%
_{c}(\tilde{p})\left( \tilde{f}_{c}(\tilde{q}_{-})+\tilde{f}_{c}(\tilde{q}%
_{+})\right) \cong 2$ for $\tilde{p}\lesssim \sqrt{\tilde{k}_{0}^{2}-\tilde{k%
}^{2}}$ the integration in Eq. (\ref{eq28}) can be done. Then we get from (%
\ref{eq28}) for $\tilde{k}_{0}-\tilde{k}\gg \hat{k}_{0}/\sqrt{d}$ after
substitution of $\varphi _{c}\left( d\right) $ from Eq. (\ref{eq23}):%
\begin{equation}
\lim_{d\rightarrow \infty }\left( \mathcal{\tilde{F}}_{\sqrt{d}\,\hat{k}}%
\left[ \tilde{f}_{c}\left( \tilde{q};d\right) ;d\right] /\sqrt{d}\right)
= \hat{F}_{0}\left( \hat{k}\right)  \label{eq31}
\end{equation}%
with the master function:%
\begin{equation}
\hat{F}_{0}\left( \hat{k}\right) \cong \left\{ 
\begin{array}{cc}
a\pi ^{-1}\hat{k}^{-2}\sqrt{\hat{k}_{0}^{2}-\hat{k}^{2}} & ,~\hat{k}\leq 
\hat{k}_{0} \\ 
0 & ,~\hat{k}>\hat{k}_{0}%
\end{array}%
\right. .  \label{eq32}
\end{equation}%
Figure \ref{fig7} presents the numerically exact result for $\mathcal{\tilde{%
F}}_{\sqrt{d}\,\hat{k}}\left[ \tilde{f}_{c}\left( \tilde{q};d\right) ;d%
\right] /\sqrt{d}$ as function of $\hat{k}=k\sigma /d^{3/2}$.

\begin{figure}[h]
\centerline{%
\includegraphics[width=8.5cm,clip=true]{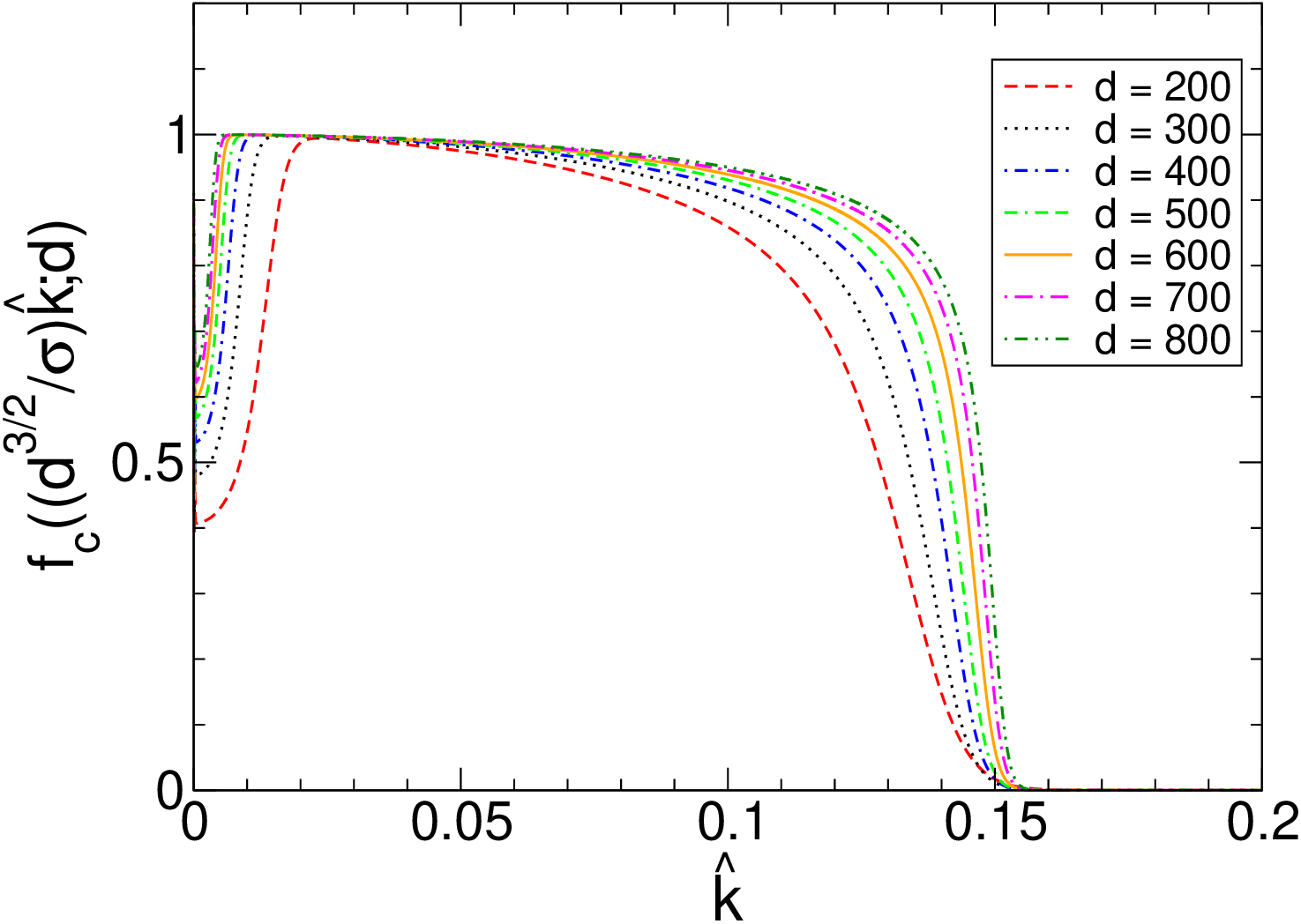}}
\caption{(colour online) The numerical critical nonergodicity parameters on the scale $\hat{k}=k%
\protect\sigma /d^{3/2}$ for various dimensions $d$.}
\label{fig8}
\end{figure}

This figure demonstrates the convergence of $\mathcal{\tilde{F}}_{\sqrt{d}\,%
\hat{k}}/\sqrt{d}$ at the glass transition singularity to the master
function $\hat{F}_{0}(\hat{k})$. For $\hat{k}<\hat{k}_{0}$, i.e. $\tilde{k}<%
\tilde{k}_{0}$ the critical nonergodicity parameters $\tilde{f}_{c}(\tilde{k};d)$ are close to one
for $d\rightarrow \infty $. Making use of Eqs. (\ref{eq22}) and (\ref{eq31})
the $\tilde{k}$- and $d$-dependence of $\tilde{f}_{c}(\tilde{k};d)$ can be
expressed as follows%
\begin{equation}
\tilde{f}_{c}\left( \tilde{k};d\right) \cong \frac{\sqrt{d}\hat{F}_{0}\left( 
\tilde{k}/\sqrt{d}\right) }{1+\sqrt{d}\hat{F}_{0}\left( \tilde{k}/\sqrt{d}\right) }
\label{eq33}
\end{equation}%
i.e. on the scale $\hat{k}=k\sigma /d^{3/2}$ it is:%
\begin{equation}
\lim_{d\rightarrow \infty }f_{c}\left( (d^{3/2}/\sigma )\hat{k};d\right) \equiv%
\hat{f}_{c}\left( \hat{k}\right) =\Theta \left( \hat{k}_{0}-\hat{k}\right) .
\label{eq34}
\end{equation}%
The convergence of the critical nonergodicity parameters to a step function is demonstrated in
Figure \ref{fig8}.

\section{Summary and Conclusions}

The liquid-glass transition for hard spheres in high dimensions $d$ has been
reinvestigated in the framework of MCT. Our aim has
not been exploring the validity of the MCT approximations for $d\rightarrow \infty$
(we come back to this point below) but to take MCT as a microscopic theory
in any dimension and to check the generic MCT-bifurcation scenario ($A_2$ singularity)
and to calculate the critical packing fraction, the corresponding nonergodicity
parameters and the exponent parameter.
The direct correlation function for hard (hyper-) spheres for finite $d$ is
not known exactly. However, if $\varphi=O(\varphi_c(d))\sim d^22^{-d}$, being
exponentially smaller than $\varphi_*(d)$ at which the Kirkwood-like instability
occurs, the corrections to the leading order result (Eq.~\ref{eq1}) can be neglected for $d\rightarrow \infty$.
This offers the possibility to calculate various quantities like 
$\varphi_c(d)$, $f_c(k;d)$, $f_c^{(s)}(k;d)$ and $\lambda(d)$
from MCT for a liquid of hard spheres in high dimensions.

\subsection{Summary}
Let us first summarize our results. The numerical solution of the MCT
equations for the collective and self nonergodicity parameters up to $d=800$ reveals non-Gaussian
dependence of the critical nonergodicity parameters $f_{c}\left( k\right) $ and $f_{c}^{\left(
s\right) }\left( k\right) $ on the wavenumber $k$ (cf. Figure \ref{fig5}). Three different $k$%
-scales were found on which $f_{c}\left( k\right) $ behaves differently. For 
$k\sigma =O(\sqrt{d})$, $f_{c}\left( k\right) $ increases from $f_{c}\left(
0\right) <1$ to a maximum value close to one. $f_{c}\left( k\right) $ stays
close to one for $k\sigma =O\left( d\right) $ and finally drops to zero for $%
k\sigma \cong \hat{k}_{0}d^{3/2}$. The decrease to zero happens on the scale 
$\hat{k}=k\sigma /d^{3/2}$ in an interval around $\hat{k}_{0}$ with width of
order $1/d$ . $f_{c}^{\left( s\right) }(k)$ and $f_{c}\left( k\right)$
are identical for $k\sigma =O(d)$ but differ for $k\sigma=O(d^{1/2})$.
The exponent parameter $\lambda(d)$ (cf. Figure \ref{fig6})
varies with $d$ even for $d>100$ and strongly exceeds the value
$\lambda (3) \cong 0.735$ \cite{FFGMS97} for $d\gg 1$.
Note that $\lambda(d)$ cannot be larger larger than one.
The critical amplitudes $h(k;d)$ \cite{GOE09} exhibit the expected $k$-dependence.
They are in anti-phase with $f_c(k;d)$, i.e. they have a maximum (minimum)
where $f_c(k;d)$ has a minimum (maximum). Particularly, on the scale $k\sigma =O(d)$
the critical amplitudes are rather small since $f_c(k;d)\approx 1$. 
The numerical results (up to $d=800$) also
show that the largest eigenvalue $E_{0}(\varphi )$ of the stability matrix 
\cite{GOE09} is not degenerate and it approaches the bifurcation point at $%
\varphi _{c}(d)$ as $(1-E_{0}(\varphi ))\sim \sqrt{(\varphi -\varphi
_{c}(d))/\varphi _{c}(d)}$, for $\varphi \rightarrow \varphi _{c}(d)$ from
above. Hence the glass transition singularity is an $A_2$ singularity,
consistent with an exponent parameter $\lambda (d)$ smaller than one, as
demonstrated by Figure \ref{fig6}. What
happens for $d=\infty $ is not clear. The
distances between the largest eigenvalues of the stability matrix obtained numerically slightly
decrease with increasing $d$. Whether or not the largest eigenvalue becomes
degenerate for $d=\infty $ is an open question, similar to the question
whether $\lambda (d)$ in Figure \ref{fig6} converges to one for $%
d\rightarrow \infty $, or not. Therefore, our numerical results do not allow to
exclude a higher order singularity at $d=\infty $.

Inspired by these numerical results we have been able to prove analytically
that the Vineyard approximation \cite{BOO80} becomes exact for $d\rightarrow \infty$
on a scale $k\sigma=O(d)$, and
that the critical packing fraction $\varphi _{c}\left( d\right) $ decays
exponentially as $2^{-d}$, with a prefactor which is quadratic in $d$. This
is consistent with the numerical result for $d\gtrsim 100$. Furthermore,
the analytical approach has also shown that the $k$- and $d$-dependence of
the critical nonergodicity parameters and of the MCT functional $\mathcal{\tilde{F}}_{\tilde{k}}[%
\tilde{f}_{c}\left( \tilde{q};d\right) ;d]$ at $\varphi _{c}\left( d\right) $
can be obtained from a master function $\hat{F}_{0}(\hat{k})$ (cf. Eqs. (\ref%
{eq31}) and (\ref{eq33})). This relationship yields%
\begin{equation}
\lim_{d\rightarrow \infty }f_{c}\left( \hat{k}\sigma ^{-1}d^{3/2};d\right)
=\Theta \left( \hat{k}_{0}-\hat{k}\right) \text{,\hspace{0.5cm}}d\rightarrow
\infty
\end{equation}%
where $\hat{k}_{0}\cong 0.15$. 

\subsection{Validity of MCT}
Although it has not been our purpose to prove or disprove the validity of MCT
for $d\rightarrow \infty$, it might be useful to comment on this question.
First of all, the vertices (\ref{eq8}) and (\ref{eq10}) seem to be exact for
$d\rightarrow \infty$, since the leading order of $c(k;d,\varphi)$ is known
analytically and the neglection of the triplet direct
correlation function $c^{\left( 3\right) }(\vec{p},\vec{q})$, which also
enters into the vertices \cite{GOE09}, is justified for $\varphi \sim
d^{2}2^{-d}$ and $d\rightarrow \infty $ (see appendix). Accordingly the
convolution approximation for the static three point correlator $\langle
\rho (\vec{k})^{\ast }\rho \left( \vec{p}\right) \rho \left( \vec{q}\right)
\rangle $ (usually done in MCT) becomes exact for $d\rightarrow \infty $ and
packing fractions such that $2^{d}\varphi (d)$ does not increase
exponentially or faster with $d$. The factorization of the static four-point
density correlator $S^{(4)}(\vec{q}_1,\vec{q}_2,\vec{q}_3,\vec{q}_4)=\frac{1}{N}
\left\langle\rho ^{\ast }\left( \vec{q}_{1}\right) \rho
^{\ast }\left( \vec{q}_{2}\right) \rho \left( \vec{q}_{3}\right) \rho \left( 
\vec{q}_{4}\right) \right\rangle $, which is needed for the projector onto
pairs of density modes, is another approximation. Similar to $S^{\left(
3\right) }(\vec{p},\vec{q})=\frac{1}{N}\langle \rho (\vec{p}+\vec{q})^{\ast
}\rho \left( \vec{p}\right) \rho \left( \vec{q}\right) \rangle $ (see
appendix) one can define a quadruplet direct correlation function $c^{\left(
4\right) }\left( \vec{q}_{1},\vec{q}_{2},\vec{q}_{3}\right) $ via a
corresponding Ornstein-Zernike equation. However, this equation is already
rather involved \cite{LEE74} such that we have not attempted to prove that
the factorization is valid for $\varphi =\varphi _{c}\left( d\right) $ and $%
d\rightarrow \infty $. So it remains an open question whether this
factorization becomes exact again. If so, the ``only'' two remaining crucial approximations
of MCT are the projection of the fluctuating force
onto pair modes and the subsequent factorization of the pair density
correlator with reduced dynamics into a product of density correlators $\phi
\left( k,t\right) $ with full dynamics. Whether these two steps become exact
for $d\rightarrow \infty $ is an interesting but also a highly nontrivial
question. Activated processes smear out the glass transition singularity.
Since it seems that the local barriers between adjacent metastable
configurations increase with increasing $d$ (see also Ref. \cite{PAR06})
these hopping processes may become suppressed for $d\rightarrow \infty $.
Even if this is true, it is not obvious that the remaining two approximation
of MCT become exact for $d\rightarrow \infty $.

One might conclude that MCT for $d\rightarrow \infty$
necessarily fails  because of $S(k;d,\varphi_c(d))\cong1$ for $k\sigma=O(d)$,
excluding the cage effect \cite{GOE09} as driving mechanism and 
one may argue that the dynamics will be described better by a
Boltzmann-Enskog equation. Indeed, a modified Enskog equation was derived
from the binary-collision expansion \cite{EF88}. But its validity for
packing fractions of order $2^{-d}$ has not been proven. 
Here, two comments are in order. \textit{First}, a cage does not necessarily
require a maximum number of adjacent spheres. For instance a (simple-) hypercubic
lattice built up of periodically arranged hyperspheres has $2d$ nearest neighbors
and a volume fraction $\varphi_{hypercube}(d) \sim d^{-d/2}$, which is much smaller
than $\varphi_c(d) \sim 2^{-d}$. Therefore, the number of contacts between neighbors
for $\varphi = \varphi_c(d)$ could be large enough in order to have a cage. Furthermore
there is evidence that the packing fraction $\varphi_{MRJ}(d)$ for maximally random
jammed states in $d$ dimensions is given by $(c_1+c_2d)2^{-d}$ \cite{SDST06}
or even with an additional term with a quadratic prefactor $c_3d^2$ \cite{P03}. These densities
are not larger than $\varphi_c(d)$. The corresponding pair correlation function
$g_2(r;d)$ flattens under an increase of $d$ from $3$ to $6$ \cite{SDST06}, i.e. comes
closer to the ideal gas value $g_{ig}(r;d)=1$, for $r > \sigma$.
\textit{Second}, concerning MCT
$S(k;d,\varphi_c(d))\cong1$ does not imply that the direct correlation function
$c(k;d,\varphi_c(d))$ and the vertices are zero. It is the quadratic dependence
of the vertices on $c(k;d,\varphi)$ in combination with its explicit $n$-dependence
and the use of the scaled variables $\tilde{k}=k\sigma/d$ which make the coupling
of the modes finite, despite the small packing fraction $\varphi_c(d)\sim d^22^{-d}$.
This is completely similar to the MCT approach to colloidal gelation for a liquid
of hard spheres with attractive Yukawa potential \cite{BF99,BFV00}. For packing
fraction $\varphi\rightarrow 0$ and potential strength $K\rightarrow\infty$ with
$K\varphi^2=\Gamma=const.$ these authors prove that $S(k)\rightarrow1$ for all $k$.
However, the vertices remain finite. At a critical value $\Gamma_c$ a liquid-gel
transition occurs. $f_c(k)$ is similar to
$f_c^{(s)}(k)$, quite analogous to our outcome. The equilibrium structure
at $\Gamma_c$ is highly ramified where the ``caging'' of a sphere is generated by
a smaller number of neighbors.

\subsection{Conclusions}
In section A. we have presented our various results from MCT in high dimensions.
Now we want to discuss the most essential findings in the light of earlier
results and will draw some conclusions.

Our critical nonergodicity parameters have a non-Gaussian $k$-dependence, in
variance with the assumption in Ref. \cite{KW87}. This discrepancy is the origin
of the different pre-exponential factor of the critical packing fraction which
we have found to be quadratic in $d$, and not linear \cite{KW87}. Due to this
quadratic $d$-dependence our MCT result for $\varphi_c(d)$
is larger than the Kauzmann packing fraction $\varphi _{K}\left(
d\right) $ (Eq. (\ref{eq3})). This cannot be true, since the packing
fraction for the Kauzmann transition (static glass transition) should be
above the packing fraction for the MCT transition (dynamical glass
transition). $\varphi _{K}\left( d\right) $ has been calculated within a
small cage expansion \cite{PAR06}, which is a kind of Gaussian
approximation. This could be the reason why $\varphi _{K}\left( d\right) $
(Eq. (\ref{eq3})) is below $\varphi _{c}\left( d\right) $ (Eq. (\ref{eq23}%
)).

As argued in subsection B. it is not necessarily true that a structureless
static correlator rules out caging. But, even if the cage effect would be absent,
the essential question would be whether the quality of both MCT
approximations \textit{necessarily} requires the existence of caging, or not.
Since an analytical investigation of the validity
of these approximations seems to be extremely difficult, a
way to get an insight is an approach by a computer simulation.
Provided such simulational results would deviate more and more from our results
with increasing dimensions this would hint at a failure of MCT for $d\rightarrow \infty$.
Such a failure would imply that MCT, which has
been interpreted as a mean field theory \cite{BIR04}, does not become exact in the limit
of high dimensions, in contrast to equilibrium phase transitions. We
hope that these concluding remarks may stimulate and encourage further
investigations, contributing to a better understanding of MCT.

\section*{ACKNOWLEDGEMENTS}

We would like to thank K. Binder, T. Franosch, M. Fuchs and W. Schirmacher for stimulating discussions.
The numerous helpful comments on our manuscript by G. Biroli, W. G\"{o}tze
and F. Zamponi are gratefully acknowledged as well.

\section*{APPENDIX}

We want to prove the following statement: for all packing fractions $\varphi
\left( d\right) $ such that $2^{d}\varphi \left( d\right) $ does not
increase exponentially or faster with $d$ for $d\rightarrow \infty $, the
static three-point correlation function $S^{\left( 3\right) }(\vec{k},\vec{k}%
^{\prime })$ reduces in the limit of high dimensions to $S\left( k\right)
S\left( k^{\prime }\right) S(|\vec{k}+\vec{k}^{\prime }|)$, i.e. the
convolution approximation becomes exact. For this proof we use the
Ornstein-Zernike equation for three-particle correlation functions \cite%
{BHP88}:%
\begin{equation}
S^{\left( 3\right) }\left( \vec{k},\vec{k}^{\prime }\right) =S\left(
k\right) S\left( k^{\prime }\right) S(|\vec{k}+\vec{k}^{\prime }|)\left(
1+n^{2}c^{\left( 3\right) }(\vec{k},\vec{k}^{\prime })\right) .
\end{equation}%
So, we have to show that $n^{2}c^{\left( 3\right) }(\vec{k},\vec{k}^{\prime
})\rightarrow 0$ for all $\vec{k},~\vec{k}^{\prime }$ for $d\rightarrow
\infty $ and $\varphi $ constrained as above. The explicit dependence of $%
c^{\left( 3\right) }(\vec{k},\vec{k}^{\prime })$ on $\vec{k},\vec{k}^{\prime
}$ does not have to be considered, as we can use for all $\vec{k},~\vec{k}%
^{\prime }$:%
\begin{eqnarray}
&&n^{2}\left\vert c^{\left( 3\right) }\left( \vec{k},\vec{k}^{\prime
}\right) \right\vert =  \notag \\
&=&n^{2}\left\vert \int d^{d}r\int d^{d}r^{\prime }~e^{-i\vec{k}\vec{r}}e^{-i%
\vec{k}^{\prime }\vec{r}^{\,\prime }}c^{\left( 3\right) }\left( \vec{r},\vec{%
r}^{\,\prime }\right) \right\vert   \notag \\
&\leq &n^{2}\int d^{d}r\int d^{d}r^{\prime }~\left\vert c^{\left( 3\right)
}\left( \vec{r},\vec{r}^{\,\prime }\right) \right\vert .  \label{app1}
\end{eqnarray}%
Now we can expand $c^{\left( 3\right) }\left( \vec{r},\vec{r}^{\,\prime
}\right) $ into diagrams, where the lines are Mayer functions and the
vertices are single particle densities. This expansion only consists of loop
diagrams. We want to show now, that the contribution of each of these
diagrams to $n^{2}\int d^{d}r\int d^{d}r^{\prime }~\left\vert c^{\left(
3\right) }\left( \vec{r},\vec{r}^{\,\prime }\right) \right\vert $ vanishes
in the limit $d\rightarrow \infty $. To do so, we apply the theorem of
Wyler, Rivier and Frisch \cite{WRH87}. This theorem states, that a loop
diagram leads to an exponentially smaller contribution to an integral like
the one appearing in the last line of Eq. (\ref{app1}), than a tree diagram
of the same order. The simplest diagram in the expansion of $c^{\left(
3\right) }\left( \vec{r},\vec{r}^{\,\prime }\right) $ reads%
\begin{equation}
\left\vert c_{0}^{\left( 3\right) }\left( \vec{r},\vec{r}^{\,\prime }\right)
\right\vert =\Theta \left( \sigma -r\right) \Theta \left( \sigma -r^{\prime
}\right) \Theta \left( \sigma -\left\vert \vec{r}-\vec{r}^{\,\prime
}\right\vert \right) ,
\end{equation}%
which can be inserted into Eq. (\ref{app1})%
\begin{eqnarray}
&&\left\vert c_{0}^{\left( 3\right) }\left( \vec{k},\vec{k}^{\prime }\right)
\right\vert \leq   \label{c3} \\
&\leq &\int d^{d}r\int d^{d}r^{\prime }~\Theta \left( \sigma -r\right)
\Theta \left( \sigma -r^{\prime }\right) \Theta \left( \sigma -\left\vert 
\vec{r}-\vec{r}^{\,\prime }\right\vert \right) .  \notag
\end{eqnarray}%
The integral occuring in Eq. (\ref{c3}) leads to an exponentially smaller
contribution than the corresponding tree diagram of the same order \cite%
{WRH87}:%
\begin{eqnarray}
&&\int d^{d}r\int d^{d}r^{\prime }~\Theta \left( \sigma -r\right) \Theta
\left( \sigma -r^{\prime }\right) \Theta \left( \sigma -\left\vert \vec{r}-%
\vec{r}^{\,\prime }\right\vert \right) \leq   \notag \\
&\leq &\alpha ^{d}\int d^{d}r\int d^{d}r^{\prime }~\Theta \left( \sigma
-r\right) \Theta \left( \sigma -r^{\prime }\right)   \notag \\
&=&\alpha ^{d}\left( V_{d}\left( \sigma \right) \right) ^{2}  \label{Wyler1}
\end{eqnarray}%
\smallskip where $V_{d}\left( \sigma \right) $ is the volume of a $d$%
-dimensional sphere with radius $\sigma $ and%
\begin{equation}
\alpha <1.  \label{alpha}
\end{equation}%
From Eqs. (\ref{c3}) and (\ref{Wyler1}) we obtain:%
\begin{equation}
n^{2}\left\vert c_{0}^{\left( 3\right) }\left( \vec{k},\vec{k}^{\prime
}\right) \right\vert \leq \alpha ^{d}\left( nV_{d}\left( \sigma \right)
\right) ^{2}.
\end{equation}%
Together with%
\begin{equation}
\varphi =nV_{d}\left( \frac{\sigma }{2}\right) =2^{-d}nV_{d}\left( \sigma
\right) 
\end{equation}%
this leads to%
\begin{equation}
n^{2}\left\vert c_{0}^{\left( 3\right) }\left( \vec{k},\vec{k}^{\prime
}\right) \right\vert \leq \alpha ^{d}\left( 2^{d}\varphi \right) ^{2},
\label{app2}
\end{equation}%
i.e. for all packing fractions, where $2^{d}\varphi $ does not increasing
exponentially or faster with $d$, we obtain from Eqs. (\ref{alpha}), (\ref%
{app2}):%
\begin{equation}
n^{2}\left\vert c_{0}^{\left( 3\right) }\left( \vec{k},\vec{k}^{\prime
}\right) \right\vert \underset{d\rightarrow \infty }{\longrightarrow }0\text{%
\hspace{0.5cm}for all }\vec{k},\vec{k}^{\prime }.
\end{equation}%
The contribution of all other diagrams are also exponentially smaller than
the corresponding tree diagrams \cite{WRH87}, which leads to:%
\begin{equation}
n^{2}\left\vert c_{i}^{\left( 3\right) }\left( \vec{k},\vec{k}^{\prime
}\right) \right\vert \leq \alpha _{i}^{d}n^{2}\left( V_{d}\left( \sigma
\right) \right) ^{2}n^{j}\left( V_{d}\left( \sigma \right) \right) ^{j}
\end{equation}%
where $j$ is the number of vertices over which it has to be integrated in
the corresponding diagram and $\alpha _{i}$ is always smaller than one. From
this we obtain%
\begin{equation}
n^{2}\left\vert c_{i}^{\left( 3\right) }\left( \vec{k},\vec{k}^{\prime
}\right) \right\vert \leq \alpha _{i}^{d}\left( 2^{d}\varphi \right) ^{j+2}
\end{equation}%
or 
\begin{equation}
n^{2}\left\vert c_{i}^{\left( 3\right) }\left( \vec{k},\vec{k}^{\prime
}\right) \right\vert \underset{d\rightarrow \infty }{\longrightarrow }0\text{%
\hspace{0.5cm}for all }\vec{k},\vec{k}^{\prime }\text{ and all }i
\end{equation}%
provided $2^{d}\varphi (d)$ does not increase exponentially or faster with $d
$.


\begin{thebibliography}{99}
\bibitem{footnote1} The result (\ref{eq19}) for $\varphi _{\ast }\left(
d\right) $ has already been obtained earlier \cite{FP87,BR88,FP99}, however,
with different numerical prefactors. The prefactor of Frisch and Percus \cite%
{FP87,FP99} $c_{0}\cong 0.871$ agrees to our prefactor, if one inserts the
correct value of the derivative of the Airy function: $A_{i}^{\prime
}(-2^{1/3}a_{0})\cong 0.701211$. The prefactor obtained by Bagchi and Rice 
\cite{BR88} is $c_{0}\cong 0.239$.
\end{thebibliography}
\end{document}